\documentclass[aps,prd,twocolumn,superscriptaddress,amsmath,amssymb,floatfix]{revtex4-2}

\usepackage{graphicx}
\usepackage{dcolumn}

\usepackage{appendix}
\usepackage[hidelinks]{hyperref}
\usepackage{caption}
\usepackage{color}
\usepackage{comment}

\begin{document}


\title{Prospects for Joint Detection of Gravitational Waves with Counterpart Gamma-Ray Bursts Detected by the HADAR Experiment}


\author{Pei-Jin Hu}
 \affiliation{Key Laboratory of Dark Matter and Space Astronomy, Purple Mountain Observatory, Chinese Academy of Sciences, Nanjing 210023, P.R.China}
 \affiliation{School of Astronomy and Space Science, University of Science and Technology of China, Hefei, Anhui 230026, P.R.China}
\author{Qi-Ling Chen}
 \affiliation{College of Physics, Sichuan University, Chengdu 610064, P.R.China}
\author{Tian-Lu Chen}
 \affiliation{The Key Laboratory of Cosmic Rays (Tibet University), Ministry of Education, Lhasa 850000, P.R.China}
\author{Ming-Ming Kang}
 \email{kangmm@ihep.ac.cn}
 \affiliation{College of Physics, Sichuan University, Chengdu 610064, P.R.China}
\author{Yi-Qing Guo}
 \email{guoyq@ihep.ac.cn}
 \affiliation{Key Laboratory of Particle Astrophysics, Institute of High Energy Physics, Chinese Academy of Sciences, Beijing 100049, P.R.China}
 \affiliation{University of Chinese Academy of Sciences, 19 A Yuquan Rd, Shijingshan District, Beijing 100049, P.R.China}
\author{Dan-Zeng Luo-Bu}
 \affiliation{The Key Laboratory of Cosmic Rays (Tibet University), Ministry of Education, Lhasa 850000, P.R.China}
\author{You-Liang Feng}
 \affiliation{The Key Laboratory of Cosmic Rays (Tibet University), Ministry of Education, Lhasa 850000, P.R.China}
\author{Qi Gao}
 \affiliation{The Key Laboratory of Cosmic Rays (Tibet University), Ministry of Education, Lhasa 850000, P.R.China}
\author{Quan-Bu Gou}
 \affiliation{Key Laboratory of Particle Astrophysics, Institute of High Energy Physics, Chinese Academy of Sciences, Beijing 100049, P.R.China}
\author{Hong-Bo Hu}
 \affiliation{Key Laboratory of Particle Astrophysics, Institute of High Energy Physics, Chinese Academy of Sciences, Beijing 100049, P.R.China}
 \affiliation{University of Chinese Academy of Sciences, 19 A Yuquan Rd, Shijingshan District, Beijing 100049, P.R.China}
\author{Hai-Jin Li}
 \affiliation{The Key Laboratory of Cosmic Rays (Tibet University), Ministry of Education, Lhasa 850000, P.R.China}
\author{Cheng Liu}
 \affiliation{Key Laboratory of Particle Astrophysics, Institute of High Energy Physics, Chinese Academy of Sciences, Beijing 100049, P.R.China}
\author{Mao-Yuan Liu}
 \affiliation{The Key Laboratory of Cosmic Rays (Tibet University), Ministry of Education, Lhasa 850000, P.R.China}
\author{Wei Liu}
 \affiliation{Key Laboratory of Particle Astrophysics, Institute of High Energy Physics, Chinese Academy of Sciences, Beijing 100049, P.R.China}
\author{Xiang-Li Qian}
 \affiliation{School of Intelligent Engineering, Shandong Management University, Jinan 250357, P.R.China}
 \affiliation{The Key Laboratory of Cosmic Rays (Tibet University), Ministry of Education, Lhasa 850000, P.R.China}
\author{Bing-Qiang Qiao}
 \affiliation{Key Laboratory of Particle Astrophysics, Institute of High Energy Physics, Chinese Academy of Sciences, Beijing 100049, P.R.China}
\author{Jing-Jing Su}
 \affiliation{College of Physics, Sichuan University, Chengdu 610064, P.R.China}
\author{Hui-Ying Sun}
 \affiliation{School of Intelligent Engineering, Shandong Management University, Jinan 250357, P.R.China}
\author{Xu Wang}
 \affiliation{School of Intelligent Engineering, Shandong Management University, Jinan 250357, P.R.China}
\author{Zhen Wang}
 \affiliation{Tsung-Dao Lee Institute, Shanghai Jiao Tong University, Shanghai 200240, P.R.China}
\author{Guang-Guang Xin}
 \affiliation{School of Physics and Technology, Wuhan University, Wuhan 430072, P.R.China}
\author{Chao-Wen Yang}
\affiliation{College of Physics, Sichuan University, Chengdu 610064, P.R.China}
\author{Yu-Hua Yao}
 \affiliation{College of Physics, Chongqing University, Chongqing 401331, P.R.China}
\author{Qiang Yuan}
 \affiliation{Key Laboratory of Dark Matter and Space Astronomy, Purple Mountain Observatory, Chinese Academy of Sciences, Nanjing 210023, P.R.China}
 \affiliation{School of Astronomy and Space Science, University of Science and Technology of China, Hefei, Anhui 230026, P.R.China}
\author{Yi Zhang}
 \affiliation{Key Laboratory of Dark Matter and Space Astronomy, Purple Mountain Observatory, Chinese Academy of Sciences, Nanjing 210023, P.R.China}
 \affiliation{School of Astronomy and Space Science, University of Science and Technology of China, Hefei, Anhui 230026, P.R.China}



\date{\today}

\begin{abstract}

    The detection of GW170817/GRB170817A implied the strong association between short gamma-ray bursts (SGRBs) and binary neutron star (BNS) mergers which produce gravitational waves (GWs). More evidence is needed to confirm the association and reveal the physical processes of BNS mergers. The upcoming High Altitude Detection of Astronomical Radiation (HADAR) experiment, excelling in a wide field of view (FOV) and a large effective area above tens of GeV, is a hope for the prompt detection of very-high-energy (VHE; > 10 GeV) SGRBs. The aim of this paper is to simulate and analyse GW/SGRB joint detections by future GW detector networks in synergy with HADAR, including the second generation LIGO, Virgo and KAGRA and the third generation ET and CE. We provide a brief introduction of the HADAR experiment for SGRB simulations and its expected SGRB detections. For GW simulations, we adopt a phenomenological model to describe GWs produced by BNS mergers and introduce the signal-noise ratios (SNRs) as detector responses. Following a theoretical analysis we compute the redshift-dependent efficiency functions of GW detector networks. We then construct the simulation of GW detection by Monte Carlo sampling. We compare the simulated results of LIGO-Virgo O2 and O3 runs with their actual detections as a check. The combination of GW and SGRB models is then discussed for joint detection, including parameter correlations, triggered SNRs and efficiency skymaps. The estimated joint detection rates are $0.09-2.52\ \mathrm{yr^{-1}}$ for LHVK network with HADAR under different possible configurations, and approximately $0.27-7.89\ \mathrm{yr^{-1}}$ for ET+CE network with HADAR.

\end{abstract}


\maketitle

\section{\label{sec:introduction}Introduction}
    GRBs are some of the most powerful explosions in the universe, lasting from 10 ms to several hours as prompt emissions and releasing most of the energy in the form of photons from keV to MeV. The durations of GRBs' prompt emissions exhibit a bimodal duration distribution, with a rough separation at $t\sim 2$ s, indicating there are two distinct groups: long GRBs (LGRBs) with $t_{\rm LGRB} > 2$ s and SGRBs with $t_{\rm SGRB} <2$ s \citep{kouveliotou1993identification, meszaros2006gamma}. LGRBs are likely produced by the collapse of massive star cores \citep{woosley1993gamma, macfadyen1999collapsars}, while SGRBs are thought to originate from the coalescence of compact binary systems, such as binary neutron stars (BNS) and neutron star-black hole (NS-BH) systems \citep{blinnikov1984exploding, paczynski1986gamma, eichler1989nucleosynthesis}, which are predicted to be accompanied by GW signals produced by compact binary system mergers.

    Evidence for SGRB progenitors was provided in 2017. The first detection of a GW signal from two coalescing neutron stars, GW~170817 \cite{abbott2017gw170817} observed by Advanced LIGO \cite{LIGO2015} and Virgo \cite{Virgo2015}, was accompanied by the coincident detection of a faint SGRB signal, GRB~170817A, within 1.7 s by Fermi \cite{goldstein2017ordinary} and INTEGRAL \cite{savchenko2017integral}. This finding strongly suggests that BNS mergers are the origin, or at least one of the origins of SGRBs \cite{abbott2017gravitational}. This finding has opened a new window for the multi-messenger astronomy, enabling the exploration of the physics of BNS merger processes with GWs and SGRBs as electromagnetic (EM) counterparts \citep{abbott2017gw170817, abbott2017gravitational}. Moreover, the realistic jets of SGRBs produced by BNS mergers provide opportunities to better understand the properties of the jet structure \citep{hallinan2017, troja2017, mooley2018, ghirlanda2019}. 
    
    Unfortunately, GRB170817A was observed off-axis from a wide angle \citep{evans2017swift, resmi2018low, d2018evolution, margutti2018binary, troja2018outflow}, resulting in a very weak signal. The next BNS-induced GW signal, GW190425 was observed nearly two years later during the O3 run of LIGO-Virgo network \cite{abbott2020gw190425}. However, its EM counterpart has not been confidently discovered. This might be partly due to the large localization region of the detection (about 8000 $\rm{deg^2}$) and the high distance, which are limitations from the detectors. The alternative reason could be its intrinsically low luminosity \cite{pozanenko2019observation, hosseinzadeh2019follow, kyutoku2020possibility}.

    To achieve such goals, more GW signals from BNS mergers with better localization are needed in observations. The coming upgrades of the second-generation LIGO, Virgo and KAGRA \citep{somiya2012detector, aso2013interferometer} will work together as the LIGO-Virgo-KAGRA network to help detect GWs from higher distances, and improve localization accuracy \cite{abbott2020prospects}. At the same time, the next generation of GW detectors designed with superior sensitivities, the Einstein Telescope (ET) \citep{punturo2010einstein, maggiore2020science} and the Cosmic Explorer (CE) \citep{abbott2017exploring, reitze2019cosmic}, are given hope to for much more distant observations. However, space experiments such as LISA \citep{danzmann1997lisa, amaro2017laser}, Taiji \cite{hu2017taiji}, and Tianqin \cite{luo2016tianqin}, are designed with much longer arm lengths, targeting low frequency GWs ranging from approximately $10^{-4}$ Hz to $1$ Hz \cite{gong2021concepts}, which are produced decades to hundreds of years before the BNS merge, according to the relationship between inspiral frequency and merger time of such binary systems \citep{cutler1994gravitational, wang2019science}. Since SGRBs are thought to emit immediately after the merger, these space experiments might not be very suitable for GW/SGRB joint detection, but they can work well as pre-announcers to find BNS systems in early stages.

    We also seek SGRB detections of prompt emissions in the VHE bands, which are urgently needed for understanding relevant topics, especially the jet structure and BNS merger processes. However, such a detection has not been achieved yet. Our hope relies on future detectors with better sensitvities, wider fields of view and larger effective areas. One powerful candidate could be the Cherenkov Telescope Array (CTA) \cite{acharya2017science}. The CTA is designed to consist of a number of telescopes with different sizes. Taking advantage of the fact that GW signals arrive sufficiently earlier than their EM counterparts which emit right after the BNS merges, the CTA can slew to the directions of the signals beforehand if the forerunner GW signals are detected by ET or CE, which can efficiently detect EM counterparts \cite{banerjee2023pre}. 
    
    Another effective approach to enhance EM counterpart observations is a wider field of view (FOV). An upcoming project ultilizing water Cherenkov lenses, the High Altitude Detection of Astronomical Radiation (HADAR) experiment, excels in a wide FOV as well as large effective area above tens of GeV. It is aimed at the prompt detection of GRBs in VHE range and relevant topics \citep{xin2021prospects, qian2022prospective}. Due to its large effective area and high sensitivities, HADAR is also capable of high distance detections. These strengths make HADAR competitively potential for the detection of EM counterparts of GWs. 
    
    To observe GW/SGRB counterparts, it is essential to coordinate GW detectors and SGRB detectors for joint detections. Our previous study \cite{ChenQL2023} worked on the potentials of HADAR to detect VHE SGRBs during the prompt phase, which has not yet been achieved by current instruments. Given the prediction of around one detection per year, it is worthwhile to further explore the potential for GW/SGRB joint detections involving HADAR. In this study, the perspective joint detection rates of GW/SGRB are calculated using different methods to analyse the joint detection capability of HADAR.

    Our work in this paper is organised as follows. In Section~\ref{sec:HADAR} We provide a brief introduction to the HADAR experiment and its expectation for prompt SGRBs based on our previous work, including the models used for SGRB detections. Section~\ref{sec:GW} focus on the detection of GWs from BNS mergers by different detector networks and their configurations. We establish the models of BNS-merger induced GW signals and the detector responses in Section~\ref{sec:GW_SNR}. Subsequently, we adopt a theoretical method for GW detection in Section~\ref{sec:GW_theo} and construct a Monte Carlo simulation of the detection of GW samples in Section~\ref{sec:GW_MC}. Joint detection analyses are discussed in Section~\ref{sec:joint}. Results of the joint detection rates are presented and analysed in Section~\ref{sec:result}. Finally, in Section \ref{sec:conclusion} we provide a conclusion of our work. 

    \begin{figure*}
        \centering
        \includegraphics[width=0.8\textwidth]{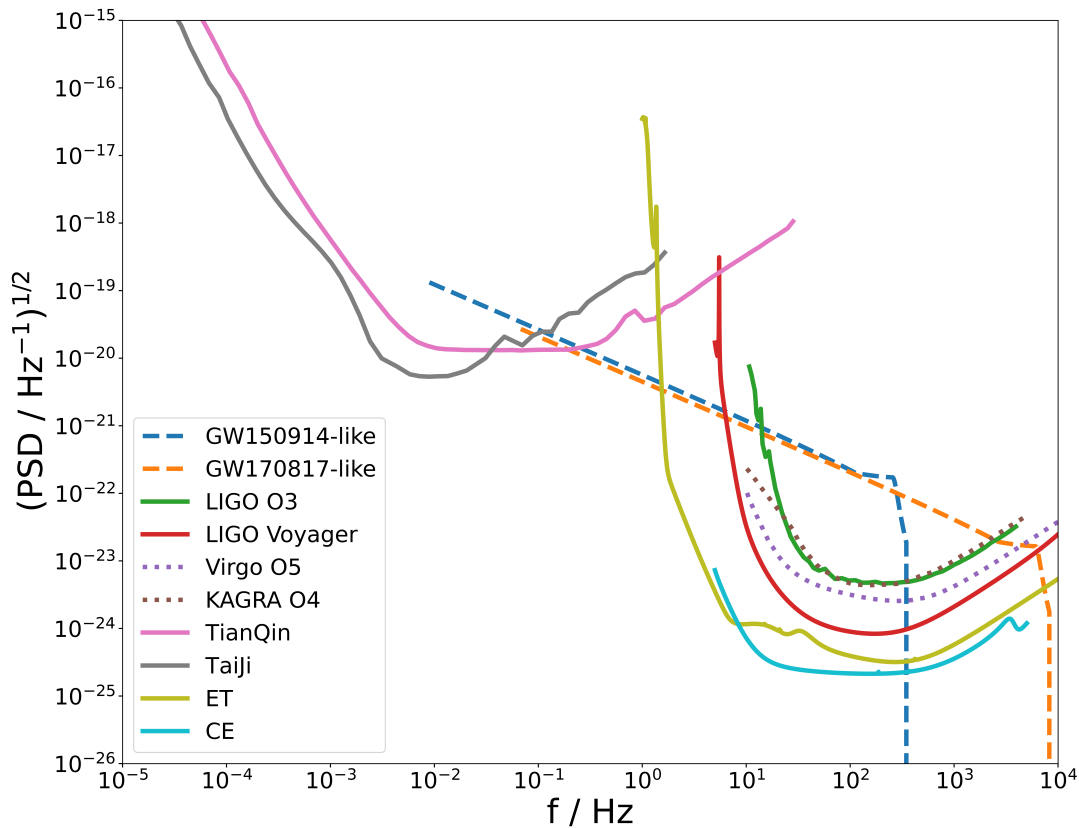}
        \caption{\label{fig:SNR}Square root of PSDs of detector noises and two types of GW signals.}
    \end{figure*}

\section{\label{sec:HADAR}HADAR Experiment and Its Expectation for Prompt SGRBs}
    The HADAR experiment is a Chinese project located in YangBaJing, Tibet of China, and is based on refractive imaging atmospheric Cherenkov telescopes (IACT), which differs from the traditional reflective IACT experiments such as H.E.S.S, CTA, etc. It consists of 4 water lenses as light collectors to refract the Cherenkov light induced by VHE comic rays and $\gamma$-rays in the atmosphere, with an angular resolution $\sim 10^{-1}$ degree and an effective area $\gtrsim 10^5\ {\rm m^2}$ above 100 GeV. For detailed descriptions of HADAR configurations and performances, references can be found in \citep{cai2017wide, chen2019performance, xin2021prospects, xin2022novel}.

    In our previous work \cite{ChenQL2023}, we developed a model to generate a set of SGRB samples and estimate HADAR's expectation for prompt emission detections. The distribution of SGRB sources, i.e. BNS merger rates $R(z)$, is constructed with a redshift dependence using the time delay model. After calculating the redshift $z$, duration $T_{90}$, energy spectra $N(E)$ and luminosity function $\Phi(L)$, we were able to generate the SGRB samples. 
    
    In this work, we carry on our previous model with appropriate adjustments. The SGRB emissions are now adapted with a jet model. The isotropically-equivalent luminosity $L_{\rm iso}$ is considered dependent on the viewing angle $\theta_{\rm v}$, which was not included in our previous work. The relation between $L_{\rm iso}$ and $\theta_{\rm v}$ is \cite{tan2020jet}:

    \begin{equation}
        L_{\rm iso}(\theta_{\rm v})=L_{\rm on}\left [  \exp \left (-\frac{\theta_{\rm v}^2}{2\theta_{\rm in}^2} \right) + C\left ( - \frac{\theta_{\rm v}^2}{2\theta_{\rm out}^2} \right ) \right ]  .
    \end{equation}
    $L_{\rm on},\ \theta_{\rm in},\ \theta_{\rm out}$ and $C$ are four free parameters of the two-Gaussian jet structure, as illustrated in Fig.~1 of \cite{tan2020jet}, where $L_{\rm on} \equiv L_{\rm iso}(0)$ is the on-axis luminosity. 
    The intrinsic luminosity function is modeled as: 
    \begin{equation}
        \Phi(L_{\rm on})=\Phi_{*}\left ( \frac{L_{\rm on}}{L_{\rm on}^{*}}\right )^{-\gamma}\exp \left ( -\frac{L_{\rm on}^{*}}{L_{\rm on}}\right ) . 
    \end{equation}
    
    Values of $\theta_{\rm in},\ \theta_{\rm out},\ C,\ L_{\rm on}^{*},\ \gamma$ are taken the same as Model A in \cite{tan2020jet}. The detection rate of prompt SGRBs does not vary much with the adaption, as to be $1.37\ {\rm yr^{-1}}$ for $\beta_{\rm ext}=-1.5$ in this adapted model.

    With the SGRB samples and the simulation of HADAR detections prepared, We will proceed to discuss the detection of GWs by different detector networks and their configurations, and develop models for the joint detection.
    

\section{\label{sec:GW}Modelling Detection of GWs}

    To simulate the detection of BNS-induced GWs, we need to build the models of GW signals from BNS systems characterized by certain properties, and calculate the detector responses to the GW signals, which are affected by several angular parameters. 

    After establishing the models to calculate GW signals, we will utilize an analytical method given in \cite{howell2019joint} to assess the GW detection performances of different detector networks by giving GW detection efficiency curves with redshift dependences. Then we will employ a Monte Carlo simulation method to obtain pseudo GW samples, allowing us to calculate relatively accurate detection rates and prepare for the simualtion of GW/SGRB joint detection.

\subsection{\label{sec:GW_SNR}The Signal-to-Noise Ratio and Detector Response}

    A GW signal is a wave of distorted space that radiates from the source, and the signals can be measured by spatial strains induced by these waves. Each GW detector has background strain noise determined by its performance and environment. The signal-to-noise ratio (SNR) of the spatial strains can be calculated by integrating over the frequency. A signal that exceeds a reasonable threshold SNR $\rho_{\rm th}$ is considered a GW signal, which is conventionally taken as $\rho_{\rm th} = 8$ \citep{abadie2010predictions, abbott2016astrophysical}. 

    We assume the standard definition for SNR given as
    \begin{equation}
        \rho = 2 \left[\ \int_{0}^{\infty} \frac{\left|\tilde{h}(f;m_1,m_2,s_1,s_2,\omega,z)\right|^{2}}{S_{n}(f)}\,\mathrm{d}f \ \right]^{1/2} , 
        \label{rho}
    \end{equation}
    where $S_n(f)$ is the power spectral density (PSD) of the detector noise. And $\tilde{h}(f)$ is the gravitational waveform from a BNS merger in the frequency domain, using the IMR spinning, nonprecessing phenomenological waveform model from \cite{ajith2011inspiral}, in the form $\tilde{h}(f) \equiv A(f) e^{-i \Psi(f)}$, where the amplitude
    \begin{equation}
        A(f)= \omega \mathcal{A}
        \begin{cases}
            f'^{-7/6}(1+\Sigma_{i=2}^3\alpha_iv^i)       & \mathrm{if} \quad f < f_1 , \\
            w_m f'^{-2/3}(1+\Sigma_{i=1}^2\epsilon_iv^i) & \mathrm{if} \quad f_1 \le f < f_2 , \\
            w_r \mathcal{L}(f,f_2,\sigma)                & \mathrm{if} \quad f_2 \le f < f_3 . 
        \end{cases}
        \label{amplitude}
    \end{equation}

    This formula is written in geometrical units(G=1, c=1). $\omega$ is a coefficient whose value depends only on the sky location and orientation, which will be explained later. $\mathcal{A}=\frac{M^{5/6}}{d_L\pi^{2/3}}\sqrt{\frac{5\eta}{24}}f_1$, where $M \equiv m_1 + m_2$ is the total mass of the binary, $d_L$ is the luminosity distance, $\eta \equiv m_1m_2/M^2$ is the symmetric mass ratio. The frequencies $f_1$, $f_2$ are the transition frequencies between the inspiral-merger and merger-ringdown stages. Above the cutoff frequency $f_3$ the amplitude is assumed negligible. $w_m,\ w_r$ are normalization constants to make the amplitude continuous over $f_1$ and $f_2$. $f'\equiv f/f_1$ and $v \equiv (\pi M f)^{1/3}$. $\mathcal{L}(f,f_2,\sigma)$ a Lorentzian function with width $\sigma$ centered around $f_2$. The phenomenological parameters $\{f_1,\ f_2,\ f_3,\ \sigma\}$ are dependent on the physical parameters $\{M,\ \eta,\ \chi\}$ and are tabulated in Table $\mathrm{\uppercase\expandafter{\romannumeral1}}$ of \cite{ajith2011inspiral}, where the single spin parameter $\chi \equiv (1+\delta)\chi_1/2 + (1-\delta)\chi_2/2$, $\delta \equiv (m_1-m_2)/M$, $\chi_i \equiv s_i/m_i^2$, $s_i$ the spin of the $i$th object. Phenomenological parameters $\{\epsilon_1,\ \epsilon_2\}$ for the merger stage and $\{\alpha_2,\ \alpha_3\}$ for the inspiral stage are dependent on $\eta$ and $\chi$, as provided in \cite{ajith2011inspiral}. 

    In this formula, the signal amplitude relies on the variables $M$, $\eta$ and $\chi$ which are intrinsic physical parameters of its BNS source; $d_L(z)$ the luminosity distance and $\omega$ the projection parameter reflect how spatial properties like distances and angular parameters from source to detector affect the response.

    Fig.~\ref{fig:SNR} displays the square root of PSDs of different detector noises, as well as two types of GW signals. The sources of these noise curves are listed in Appendix~\ref{apd:noisecurve}. Space-borne detectors TianQin and TaiJi are more sensitive in the frequency band $10^{-5} - 10^1$ Hz, while ground-based instruments are better at bands from $10^1$ to $10^4$ Hz. The sensitive bands also limit the integral ranges of detected SNRs. The dashed lines are calculated from simulations of GW170817 and GW150914, produced from a BNS merger and a BBH merger, respectively. The parameters used are $m_1=1.48M_{\odot},\ m_2=1.26M_{\odot}$ at 40 Mpc for GW170817 and $m_1=35.3M_{\odot},\ m_2=29.6M_{\odot}$ at 440 Mpc for GW150914. 

    The parameter $\omega$ mentioned in Eq.~\ref{amplitude}, described as the projection parameter (see Appendix A in \cite{dominik2015double}), represents the response of a single detector to the angular distribution of power for a GW source dominated by $(l, |m|)=(2,2)$ over the two-dimensional sky location $(\theta, \phi)$, inclination $\iota$, and polarization $\psi$. This parameter is first summarized by \cite{finn1993observing}. The cumulative distribution $C(\omega)$ can be calculated analytically \cite{finn1996binary}, and Dominik provided a numerical result in his passage \cite{dominik2015double}, which was expanded to multiple detectors.

    The formula of $\omega$ is given as
    \begin{equation}
        \omega=\frac{1}{2}\sqrt{(1+\cos^2\iota)^2F_+^2+4\cos^2\iota F_{\times}^2} ,
    \end{equation}
    where $F_+,F_\times$ are the detector's antenna-pattern functions \cite{finn1993observing}, and $\iota$ is the inclination between the face-on direction of the binary source (where the GW signal is strongest) and the direction from the source to the observer. We build the functions using Schutz's model \cite{schutz2011networks}, who transformed the functions for each detector into unified ones in the geographic coordinate system of the Earth, expressed using each detector's geographic location and orientation. Thus, we can express the patterns in multi-detector networks. 

    The expressions are given as
    \begin{eqnarray}
        F_+=\sin\eta[a\cos(2\psi)+b\sin(2\psi)] ,
        \\
        F_\times=\sin\eta[b\cos(2\psi)-a\sin(2\psi)] ,
    \end{eqnarray}
    where 
    \begin{eqnarray}
        a=&& \frac{1}{16}\sin(2\chi)[3-\cos(2\beta)][3-\cos(2\theta)]\cos[2(\phi-\lambda)] \nonumber\\
        &&+\frac{1}{4}\cos(2\chi)\sin(\beta)[3-\cos(2\theta)]\sin[2(\phi-\lambda)] \nonumber\\
        &&+\frac{1}{4}\sin(2\chi)\sin(2\beta)\sin(2\theta)\cos(\phi-\lambda) \nonumber\\
        &&+\frac{1}{2}\cos(2\chi)\cos(\beta)\sin(2\theta)\sin(\phi-\lambda) \nonumber\\
        &&+\frac{3}{4}\sin(2\chi)\cos^2(\beta)\cos^2(\theta) ,
    \end{eqnarray}
    
    \begin{eqnarray}
        b=&& \cos(2\chi)\sin(\beta)\sin(\theta)\cos[2(\phi-\lambda)] \nonumber\\
        &&-\frac{1}{4}\sin(2\chi)[3-\cos(2\beta)]\sin(\theta)\sin[2(\phi-\lambda)] \nonumber\\
        &&+\cos(2\chi)\cos(\beta)\cos(\theta)\cos(\phi-\lambda) \nonumber\\
        &&-\frac{1}{2}\sin(2\chi)\sin(2\beta)\cos(\theta)\sin(\phi-\lambda) ,
    \end{eqnarray}

    \begin{table*}
        \caption{\label{tab:ground}Configurations of LIGO, Virgo and KAGRA}
        \begin{ruledtabular}
 	\begin{tabular}{cccccc}
 		name & Label &  $\eta$ &  $\chi$  & $\beta$ &  $\lambda$ \\[0.5ex] 
            \hline
 		LIGO Livingston, LA&L & $90^{\circ}$& $242^{\circ}$&
 		$90^{\circ} 46^{\prime} 27.3^{\prime\prime} \mathrm{W}$ & $30^{\circ}33^{\prime}46.4^{\prime\prime} \mathrm{N}$    \\ 
 		LIGO Hanford, WA& H & $90^{\circ}$ & $171^{\circ}$ & $119^{\circ}24^{\prime}27.6^{\prime\prime} \mathrm{W}$& $46^{\circ}27^{\prime}18.5^{\prime\prime} \mathrm{N}$   \\ 
 		Virgo, Italy & V& $90^{\circ}$ & $116.5^{\circ}$ &$10^{\circ}30^{\prime}16^{\prime\prime} \mathrm{E}$ & $43^{\circ}37^{\prime}53^{\prime\prime} \mathrm{N}$ \\ 
 		KAGRA, Japan & K&  $90^{\circ}$ &$70^{\circ}$ & $137^{\circ}10^{\prime}48^{\prime\prime} \mathrm{E}$ &$36^{\circ}15^{\prime}0^{\prime\prime} \mathrm{N}$ \\
 	\end{tabular}
        \end{ruledtabular}
    \end{table*}

    Here, $\eta$ is the angle between the interferometer arms(opening angle). $\psi$ is the wave polarization angle. $\chi$ is measured counterclockwise from East to the bisector of the interferometer arms. $\beta$ and $\lambda$ are the latitude and the longitude of the detector. $\theta$ and $\phi$ are the latitude and the longitude of source. The values of $\eta,\ \chi,\ \beta,\ \lambda$ of LIGO, Virgo and KAGRA are tabulated in Table~\ref{tab:ground}.

    The aim of this transformation is to calculate the network SNR in consistent coordinates. For a network with N detectors, the network SNR is defined as \cite{finn2001aperture, schutz2011networks}: 
    \begin{equation}\label{rho_net}
        \rho_{\rm net}^2 = \sum_{k = 1}^{N} \rho_k^2 . 
    \end{equation}
    
    For ground instruments observing the same source, their SNRs are correlated due to fixed relative positions and orientations, which determines the network SNR. With a certain spatial distribution of sources, different layouts of detectors lead to different distributions of network SNRs. This corresponds to distributions of the network response $\omega_{\rm net}$, as shown in Appendix~\ref{apd:omgdis}.

    There are some assumptions made in this work. For the waveform model of the signals, only nonspinning ones $(s_1,\ s_2=0)$ are considered. When it comes to theoretical analysis, all detectors of the same network are assumed to have equal sensitivities for convenience. Following the convention, $\rho_{\rm th}$ is set to 8 for the single detector, and as a proxy for detector networks as well.

\subsection{\label{sec:GW_theo}Theoretical Analysis of GW Detection}
    We adopt a theoretical method from \cite{howell2019joint} which models the detection efficiency function and calculates corresponding detection rates. 

    Following \cite{howell2019joint}, we can build the detection rate calculation with a redshift dependence by
    \begin{equation}
        R_{\rm GW}(z)=R_{\rm BNS}(z)\Sigma_{\rm GW}(z)\Omega_{\rm GW} ,
    \end{equation}
    where $R_{\rm GW}(z)$ is the detection rate of GW relying on redshift $z$, $R_{\rm BNS}(z)$ is the redshift distribution of BNS merger rate. $\Omega_{\rm GW}$ is the duty cycle of a typical observing run, and $\Sigma_{\rm GW}(z)$ is the detector efficiency function, mathematically the ratio of the fraction that could be detected of all sources at redshift $z$. 

    The GW efficiency function $\Sigma_{\rm GW}(z)$ can be calculated by mapping it to the cumulative probability distribution of the projection parameter $C(\omega)$. The optimal SNR for any source $\rho_{\rm opt}$ can be calculated from Eq.~\eqref{rho} in the particular case $\omega = 1$, the best case in which it is a face-on, overhead source. And the SNR of the detector response from the same source is $\rho=\omega \rho_{\rm opt}$. In critical situations when signals are of the least strength, $\omega$ gets $\omega_{\rm min}=\rho_{\rm th}/\rho_{\rm opt}$, the minimum value for a specific source $\rho_{\rm opt}$ to be detected as a signal. Since $\omega$ is a constant value for $\rho_{\rm opt}$, it can be written in the redshift-dependent form $\rho_{\rm opt}(z)$. Thus the efficiency function is constructed as
    \begin{equation}
        \Sigma_{\rm GW}(z)=C(\omega \ge \omega_{\rm min})=C(\omega_{\rm min})=C(\rho_{\rm th}/\rho_{\rm opt}(z))
    \end{equation}

    There is an analytical approximation to the cumulative distribution $C(\omega)$ of any single detector \cite{finn1996binary}, while $C(\omega)$ of detector networks can be calculated numerically via Monte Carlo sampling on the angular parameters of GW signals, see Appendix~\ref{apd:omgdis}, where the numerical results are displayed with cumulative curves, including a single detector, LH, LHV and future LHVK network.

    \begin{figure}[b]
        \centering
        \includegraphics[width=0.5\textwidth]{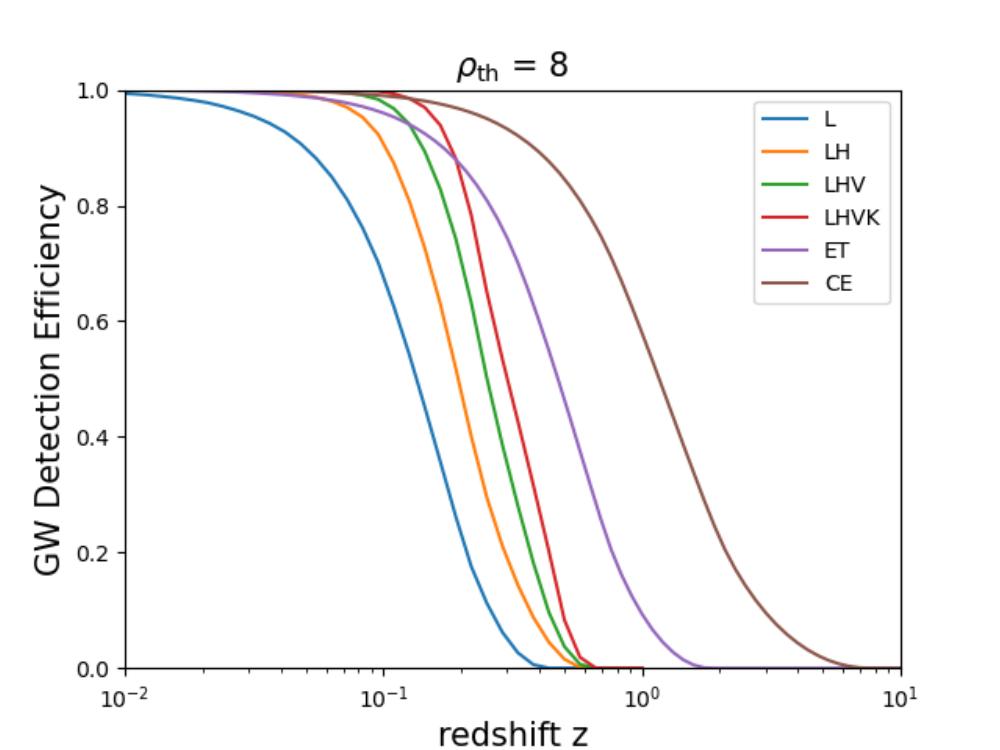}
        \caption{\label{fig:GWeff}GW efficiency functions based on different detector configurations. }
    \end{figure}

    With the specific $C(\omega)$, the GW detection efficiency can be calculated, as shown in Fig.~\ref{fig:GWeff}. Here we adopt the Voyager sensitivity curve as a standard for detectors L, H, V, K in cooperation. For ET and CE we adopt their respective sensitivities. Fig.~\ref{fig:GWeff} suggests that horizons of the LIGO-Virgo-KAGRA cooperation are limited to around $z \sim 0.6$. However, with the future generation ET and CE it's possible to expand horizons beyond that, with CE's horizon up to $z \sim 5$.

    \begin{table*}[t]
        \caption{\label{tab:GWrate}GW observations and program simulations during O2 and O3 run. %
        }
        \begin{ruledtabular}
        \begin{tabular}{ccccc}
        \textrm{Observing Run}& 
        \textrm{Equivalent Observing Time(days)}&
        \textrm{Observation(s)}&
        \textrm{Method A}&
        \textrm{Method B}\\
        \hline 
        O2 & 120.6 & 1 & $0.69^{+1.43}_{-0.55}$ & $0.72^{+1.50}_{-0.57}$\\
        O3 & 156.4 & 1 & $4.0^{+8.3}_{-3.2}$ & $2.7^{+5.6}_{-2.1}$\\

        \end{tabular}
        \end{ruledtabular}
        \end{table*}

\subsection{\label{sec:GW_MC}Monte Carlo Sampling of GW Detection}
    In this section, we adopt a simulation method based on Monte Carlo sampling. A set of parameters $\{m_1,\ m_2,\\ s_1,\ s_2,\ z,\ \theta,\ \phi,\ \psi,\ \iota \}$ are generated to describe a GW signal from its BNS source. Masses $m_1$ and $m_2$ are the individual masses of the two stars, which follow the neutron star mass distribution from \cite{kiziltan2013neutron}: 
    \begin{equation}
        P_{NS}(m) = 2\phi \left(\frac{m-\mu}{\sigma}\right) \Phi \left(  \frac{(m-\mu)\alpha}{\sigma}\right)
    \end{equation}
    where $\phi(x)$ and $\Phi(x)$ are the standard normal density and cumulative density functions. For BNS systems, \cite{kiziltan2013neutron} found the phenomenological parameters $\mu =1.33$, $\sigma = 0.11$ and $\alpha = -0.03$. We conservatively assume that the two NS masses of the binary are independent. Their individual spins are set $s_1=s_2=0$. The redshift z follows the distribution of BNS merger rate $R(z)$, which is mentioned in the SGRB model \cite{ChenQL2023}. $(\theta,\ \phi)$ is the sky position of the BNS generated by a random 3-dimensional vector. $\psi$ is the polarization of the GW, generated by a uniform random angle. The inclination angle $\iota$ follows a sine distribution $P(\iota)\propto \sin(\iota)$. According to Eq.~\ref{rho}, the SNR of each simulated GW can be then calculated with the detector noise.

    We tested our models under conditions of LIGO O2 and LIGO-Virgo O3. For O2 we consider the LH network to be observing together, with an observing period of 268 days from November 30, 2016 to August 25, 2017 and a network duty cycle $45\%$ \cite{GWTC1-2019}. For O3 we consider the LHV network and a 3-det network operation time equivalent to 156.4 days \cite{GWTC2-2021, GWTC3-2021}. As for BNS merger rates, we follow the model we used in our previous work \cite{ChenQL2023}, which is based on \cite{abbott2017gw170817}. 
    
    In real observations, there was one BNS-merger GW detected during O2 run (GW170817) and one during O3 run (GW190425). Table~\ref{tab:GWrate} lists the results of our simulations in comparison with real observations. The column Method A refers to the method in Section \ref{sec:GW_theo} using GW efficiency functions. Sensitivities used for all are L O2 and O3. Since the sensitivity of Virgo is not as good in O3, the estimate is a little overrated here. The column Method B refers to the MC simulation in this section, which uses the exact sensitivities to each detector. We can see that the results are consistent with each other.

    \begin{figure}[b]
        \centering
        \includegraphics[width=0.5\textwidth]{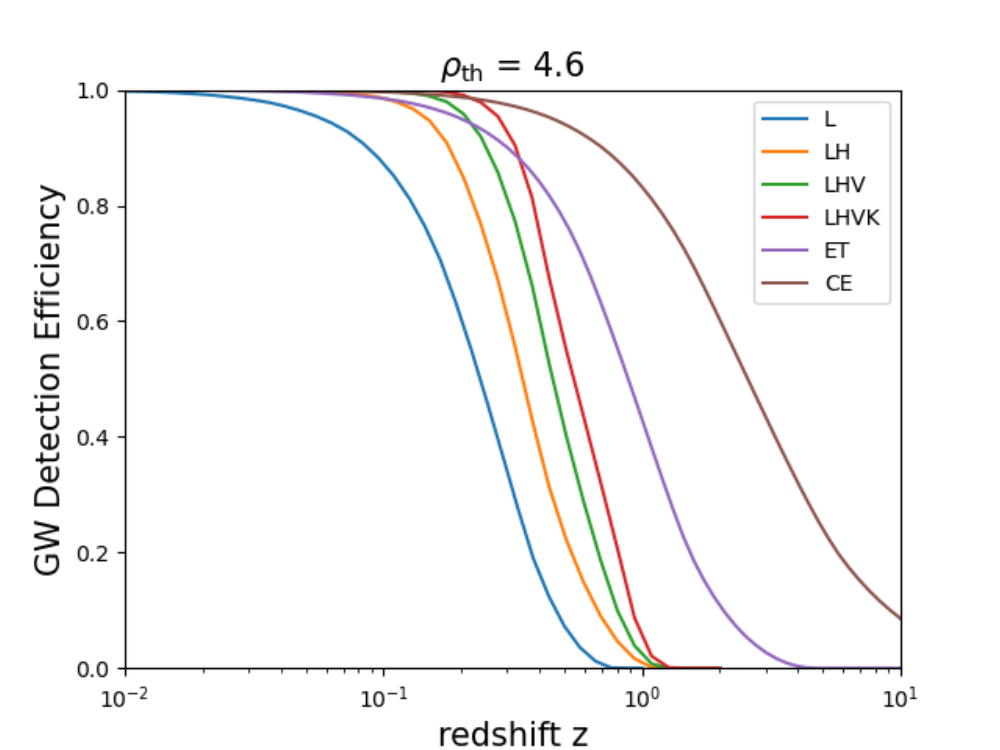}
        \caption{\label{fig:GWeff2}Triggered GW efficiency functions based on different detector configurations. }
    \end{figure}

\section{\label{sec:joint}Joint Detection of GW/SGRBs with HADAR}
    Now that the GW detection framework is constructed, the next step is to connect GW experiments in synergy with the HADAR experiment to model their joint detection. From the same BNS sources, these two events are associated in spatial and physical properties, as we will discuss later. Additionally, there is a modification of GW signal thresholds that need to be lowered due to the triggers of SGRB detections. We'll then analyze the efficiency skymaps of different detector networks together with HADAR, so as to provide a direct view of the effectiveness of the joint observations.  

    \begin{figure*}
        \includegraphics[width=0.45\textwidth]{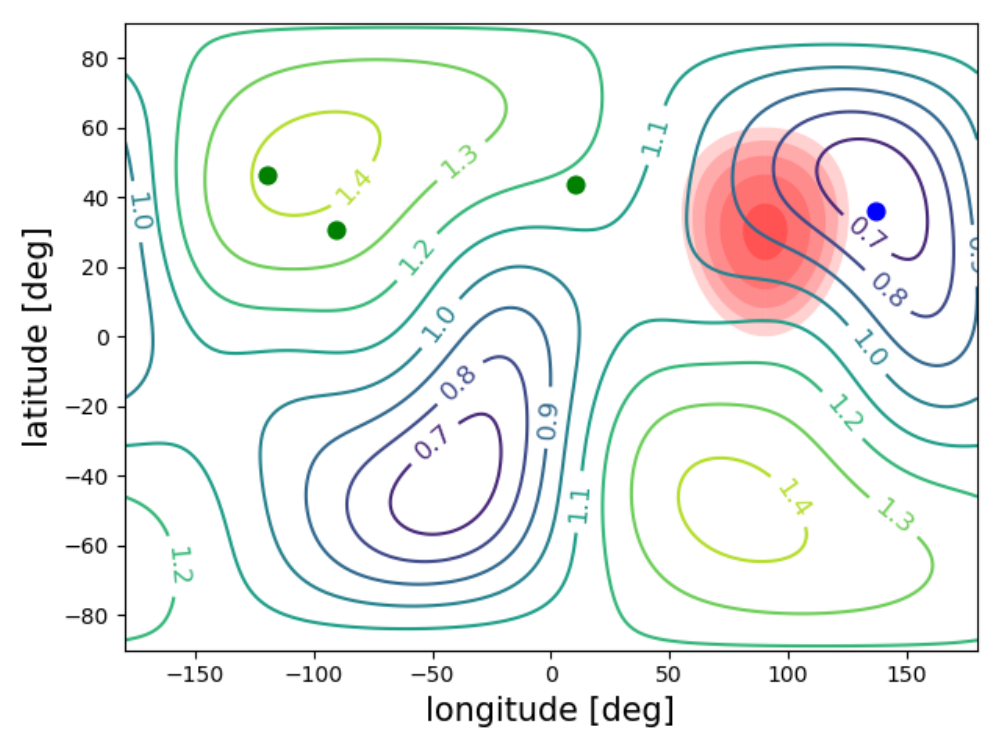}
        \includegraphics[width=0.45\textwidth]{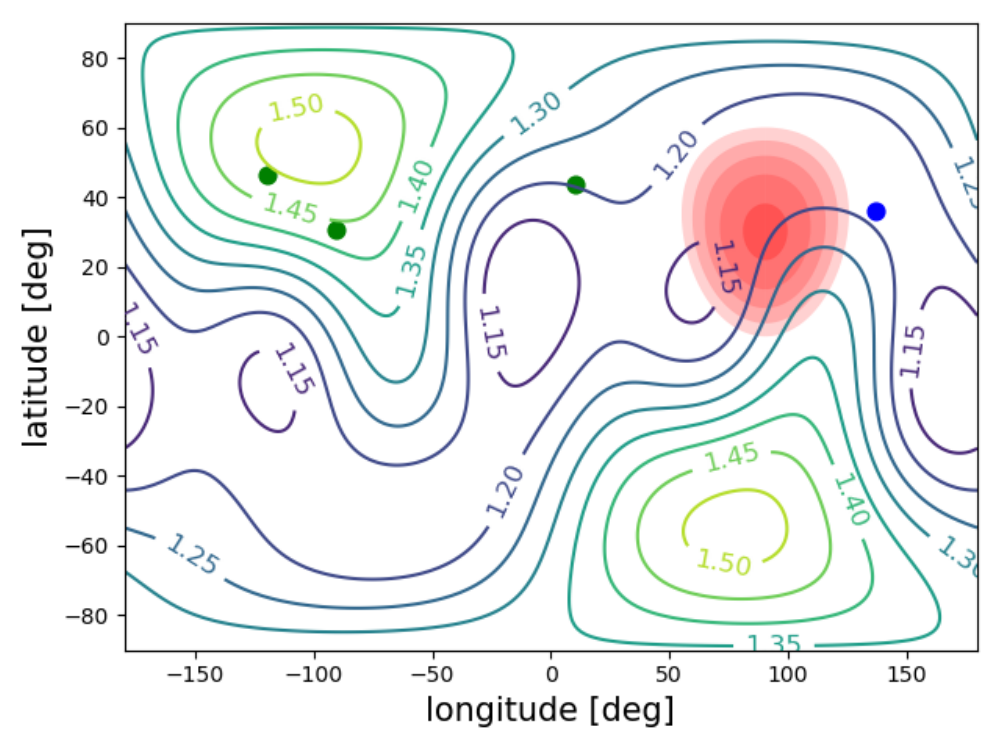}
        \caption{\label{fig:skymap1}The skymaps of joint detection. Left: HADAR + LHV; Right: HADAR + LHVK. }
    \end{figure*}

\subsection{\label{sec:joint_para}Parameter Correlations}
    In joint detection simulations, the parameter correlations between GWs and their counterpart SGRBs must be taken into consideration. 
    
    For GW parameters $\{m_1,\ m_2,\ s_1,\ s_2,\ z,\ \theta,\ \phi,\ \psi,\ \iota\}$, the inclination angle $\iota$ should be considered the same as the viewing angle of the corresponding SGRB jet. The redshift $z$ and the sky location $(\theta, \phi)$ of the two events are obviously the same since they're from the same source. Polarization $\psi$ is thought irrelevant and spins $s_1,\ s_2$ are set zero. Technically, energy emitted by the SGRB is associated with the star masses $m_1,\ m_2$ of the BNS system. In our work, they are considered independent as a simplification.

\subsection{\label{sec:joint_threshold}Thresholds of Triggered GW Detection with EM Counterparts}

    In terms of GW thresholds in joint detection, the EM counterpart detections shall be taken into consideration. For an EM-triggered GW detection, the more constrained temporal and spatial coverages reduce the false alarm rate (FAR), thus lowering the SNR threshold for the GW detection search, and increasing the detection horizon. The relation between the SNR thresholds for EM-triggered GW searches $\rho^{trig}_{\rm th}$ and untriggered ones $\rho_{\rm th}$ is constructed as \cite{bartos2015beyond, patricelli2016prospects}:
    \begin{equation}
        \rho^{trig}_{\rm th}=\sqrt{2\log\left[\exp\left(\frac{\rho_{\rm th}^{2}}{2} \right)\frac{t_{\rm obs}\times\Omega}{t_{\rm obs,0}\times\Omega_{0}}\right]}
    \end{equation}
    where $\Omega$, $\Omega_{0}$ and $t_{\rm obs}$, $t_{\rm obs,0}$ are the relative sky regions and observation durations, respectively. 

    We adopt \cite{patricelli2016prospects} to take $\Omega_{0}=40000\ \deg^2$, $\Omega = 100\ \deg^2$, $t_{obs,0} = 1\ {\rm yr}$, and $t_{\rm obs}=\delta t \times N_{\rm sGRB}$ where $\delta t$ is the GW search time window around the EM trigger which we sufficiently take to be $6s$ and $N_{\rm sGRB}$is the number of expected sGRB detections per year within the GW detector horizon, which is conservatively taken as $1\ {\rm yr^{-1}}$ for our estimations. 
    
    For $\rho_{\rm th} = 8$ we get a triggered threshold $\rho^{trig}_{\rm th} \sim 4.6$. This lowered threshold corresponds to new GW detection efficiency curves with broader horizons, which are re-calculated in Fig.~\ref{fig:GWeff2}. In such situations, LHVK reaches $z \sim 1$, ET reaches $z \sim 4$ and CE reaches beyond $z = 10$.

\subsection{\label{sec:joint_network}Joint Detection Skymaps}

    With each network of ground-based GW detectors, we can draw a skymap of the amplitude patterns, which corresponds to the geographic distribution of projection parameter $\omega$ of optimal-orientated signals from directions of all sky. Such a skymap represents the relative efficiency of the network for a signal coming from a paticular direction of the sky, and is able to give advice on the layouts of networks. For this purpose, we involve in HADAR's detection region for the joint detection. As a ground-based instrument, HADAR is located at [$30.1^{\circ}$ N, $90.5^{\circ}$ E] in Yangbajing, Tibet of China with a $30^{\circ}$ FOV. 

    Fig.~\ref{fig:skymap1} shows the comparison of HADAR + LHV network and HADAR + LHVK network. The contour lines display the amplitude patterns of LHV and LHVK, while the green spots represent positions of L, H, V and the blue one K. The red-colored region shows HADAR's FOV, the shade indicating detection efficiency of HADAR. From the comparison, we can see the operation of KAGRA can make up for the relatively blind areas of LHV network, increasing the detection efficiencies within HADAR's region. The joining of KAGRA raises the minimum $\omega$ from 0.7 to 1.15, and the maximum from 1.4 to 1.5. Note that the GW detectors are assumed with the same sensitivities.

\section{\label{sec:result}Results}

    \begin{figure*}
        \includegraphics[width=0.45\textwidth]{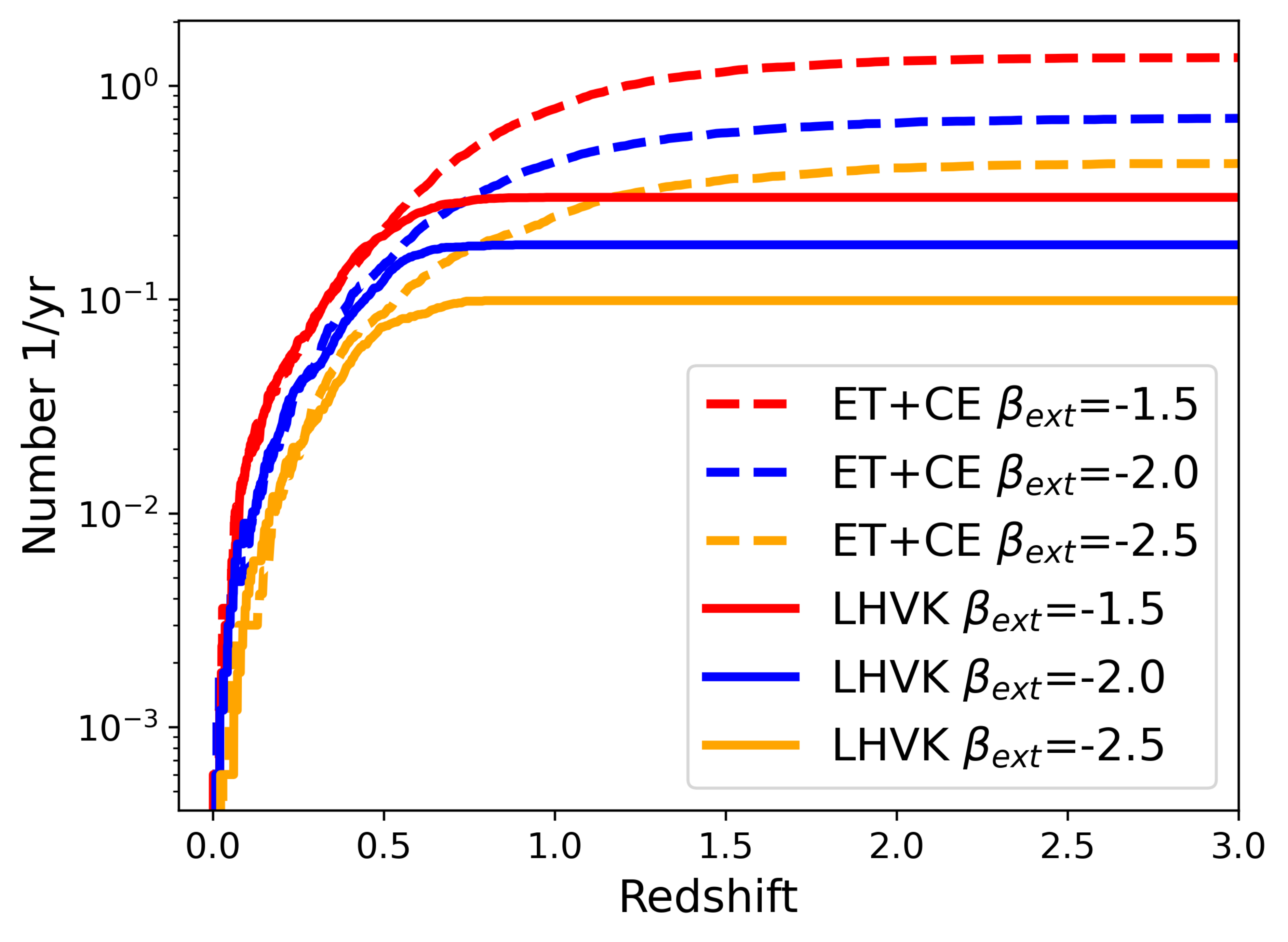}
        \includegraphics[width=0.45\textwidth]{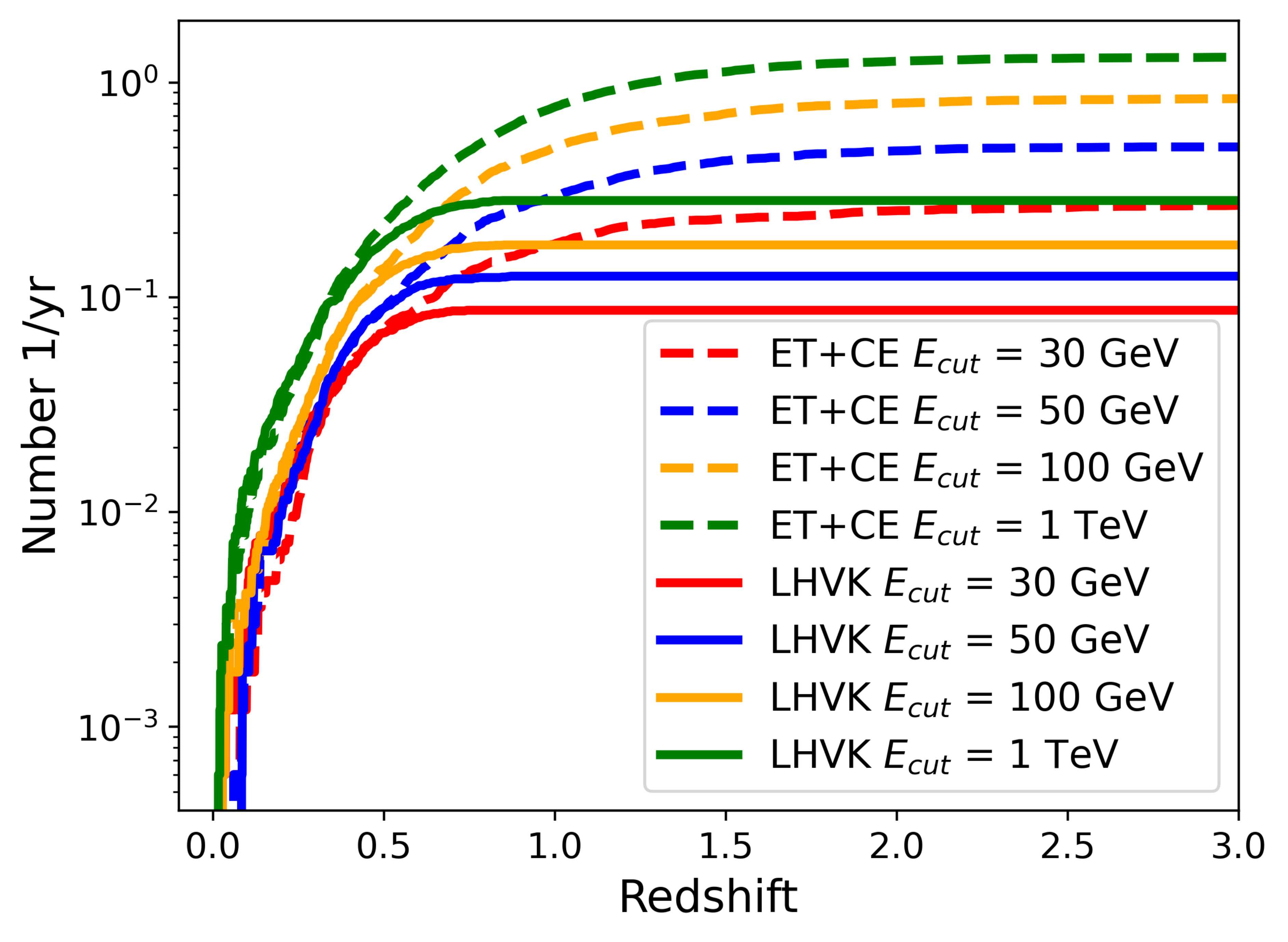}
    \caption{\label{fig:rate1}Cumulative joint detection rates under different SGRB spectral parameters. Left: different spectral indices of the extra component. Right: different spectral cut-off energies. }
    \end{figure*}

    \begin{figure}[hb]
        \includegraphics[width=0.45\textwidth]{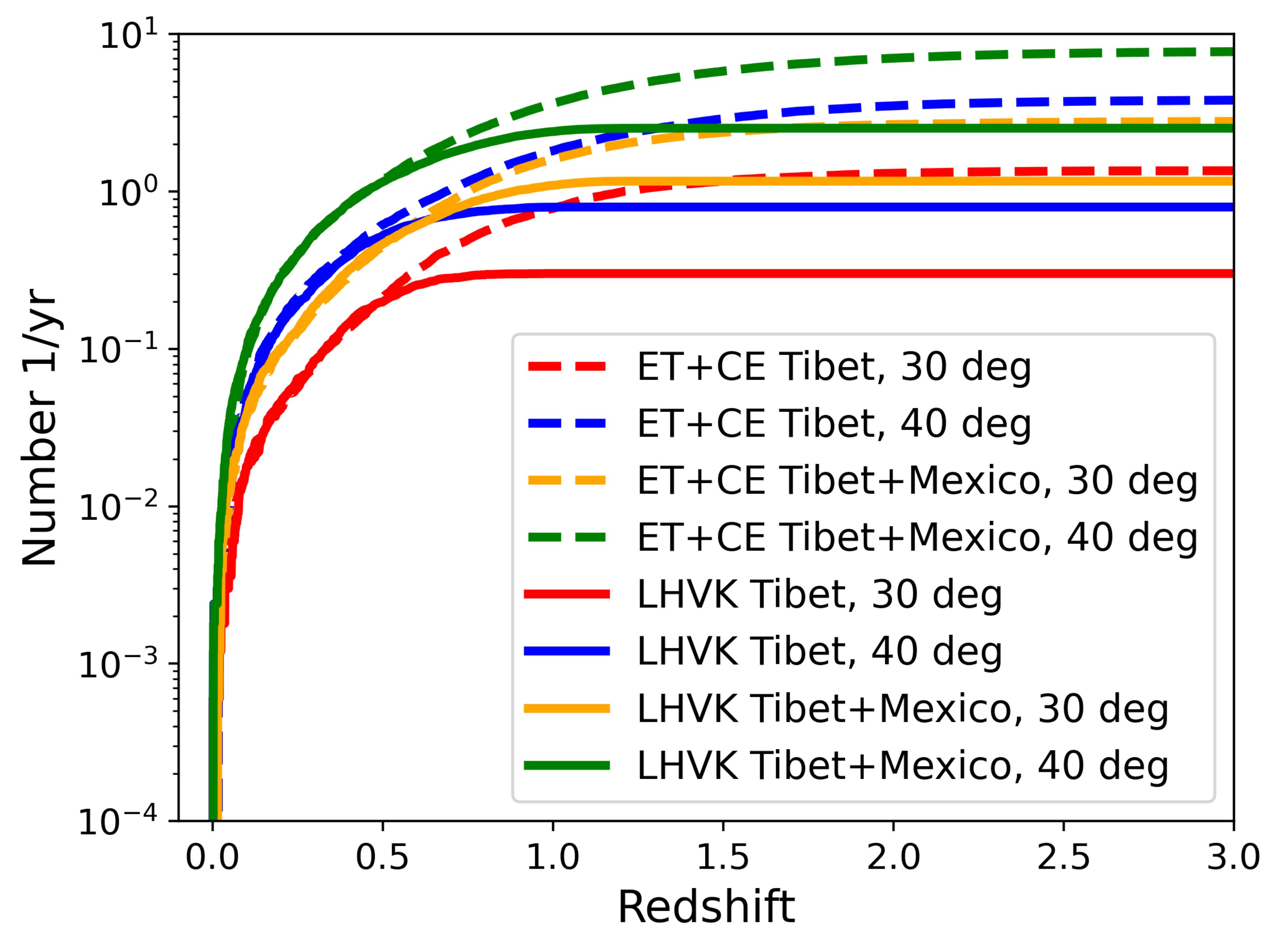}
        \caption{\label{fig:rate2}Cumulative joint detection rates with different configurations of HADAR. }
    \end{figure}

    In this section, we give the results of joint detection rates estimated by our simulations. Different configurations of HADAR and GW detectors are taken into account. Two scenarios of GW detector networks are applied, one to be the future LHVK with LIGO Voyager, Virgo O5 and KAGRA O4 sensitivities, and the other the ET + CE network with their designed sensitivities.

    The influence of the variations of SGRB spectral parameters is necessarily considered. Fig.~\ref{fig:rate1} represents the cumulative joint detection rates under different SGRB spectral parameters. The solid lines represent results with network LHVK, while dashed lines represent ET + CE. As in the left panel, different spectral indices of the extra component are simulated. Corresponding to $\beta_{\rm ext}$ = -1.5, -2.0, -2.5 (with no energy cutoff), the joint detection rates are 0.30, 0.18, 0.09 ${\rm yr^{-1}}$ with LHVK and 1.36, 0.71, 0.44 ${\rm yr^{-1}}$ with ET + CE. In the right panel are different spectral cut-off energies. When $E_{\rm cut}$ = 30 GeV, 50 GeV, 100 GeV, 1 TeV (with $\beta_{\rm ext}$ = -1.5), the joint detection rates are 0.09, 0.13, 0.18, 0.28 ${\rm yr^{-1}}$ with LHVK, and 0.27, 0.50, 0.85, 1.31 ${\rm yr^{-1}}$ with ET + CE. 
    
    Furthermore, some promotions of HADAR configurations are considered, as represented in Fig.~\ref{fig:rate2}. The basic configuration is HADAR seated in Yangbajing, Tibet of China, with the FOV of $30^{\circ}$. The estimated detection rates are 0.30 ${\rm yr^{-1}}$ with LHVK and 1.37 ${\rm yr^{-1}}$ with ET + CE. One promotion is to expand the FOV to $40^{\circ}$, which sacrifices the sensitivity and increases the background by 4 times. Another promotion is a future possibility to build a second instrument of HADAR elsewhere, which we choose to be [$19^{\circ}$ N, $97.3^{\circ}$ W] in Mexico (where HAWC resides). For the promotions of a single HADAR instrument in Tibet with $40^{\circ}$ FOV, two HADAR instruments in Tibet + Mexico with $30^{\circ}$ FOV and that with $40^{\circ}$ FOV, the joint detection rates are 0.80, 1.16, 2.52 ${\rm yr^{-1}}$ with LHVK, and 3.87, 2.80, 7.89 ${\rm yr^{-1}}$ with ET + CE, respectively.


\section{\label{sec:conclusion}Conclusions}
    Since the observation of GW170817/SGRB170817A, the association of SGRBs and BNS mergers has provided a new perspective for multi-messenger astronomy, as well as insights into the physics of BNS merger and the jet structure. While the HADAR experiment is a hope for prompt VHE SGRB detections, the improving GW detectors offer chances for joint detections of GW/SGRBs.

    Our study focuses on GW detection and joint detection by present and future GW instruments in synergy with HADAR. We have developed phenomenological models for GW signals and conducted analyses of different GW detectors, including single L, LHV, LHVK, ET and CE, evaluating their horizons and other performances. After that, we calculated joint detection rates under different SGRB parameters and detector configurations. For future LHVK, the joint rates range from $0.09-2.52\ \mathrm{yr^{-1}}$, and for ET + CE the rates increase to $0.27-7.89\ \mathrm{yr^{-1}}$, demonstrating a significant capability for joint detection. 
    
    Based on our work, we conclude that the future LHVK network, as well as ET and CE, show significant potential for joint detections with HADAR. We sincerely hope that future LHVK, HADAR, ET and CE experiments will be successfully completed and prove the joint detections consistent with our expectations. 


\begin{acknowledgments}
    This work is supported by the National Natural Science Foundation of China (Nos. 12220101003, 12321003), the Project for Young Scientists in Basic Research of Chinese Academy of Sciences (No. YSBR-061). 
\end{acknowledgments}

\appendix
\section{\label{apd:noisecurve}GW Interferometer Sensitivity Noise Curves}
 The sensitivity noise curves used in Fig.~\ref{fig:SNR} and for calculation are as follows: For the O2 and O3 run of Advanced LIGO, we extract noise curves from \cite{GWTC1-2019} and \cite{GWTC3-2021}, respectively. For the LIGO A+ design curve we use the model from \url{https://dcc.ligo.org/LIGO-T1800042/public}. For LIGO Voyager we use the data from \url{https://dcc.ligo.org/LIGO-T1500293/public}. For Virgo O5 and KAGRA O4, the data are extracted from \cite{abbott2020prospects} using the lower bounds of the noise curves. For TianQin and TaiJi, the curves are found in \cite{gong2021concepts}. For CE sensitivity we use the model from \url{https://cosmicexplorer.org/sensitivity.html}. And for ET we extract data from \url{https://www.et-gw.eu/index.php/etsensitivities} using the ET-D model.

\section{\label{apd:omgdis}Cumulative Amplitude Distribution for Detector Networks}
    Assuming all detectors of the network are identical, the amplitude projection parameter of the network can be defined as $\omega_{\rm net}^2 = \sum_{k = 1}^{N} \omega_k^2$ according to Eq.~\ref{rho_net}. 

\begin{figure}[!h]
        \includegraphics[width=0.45\textwidth]{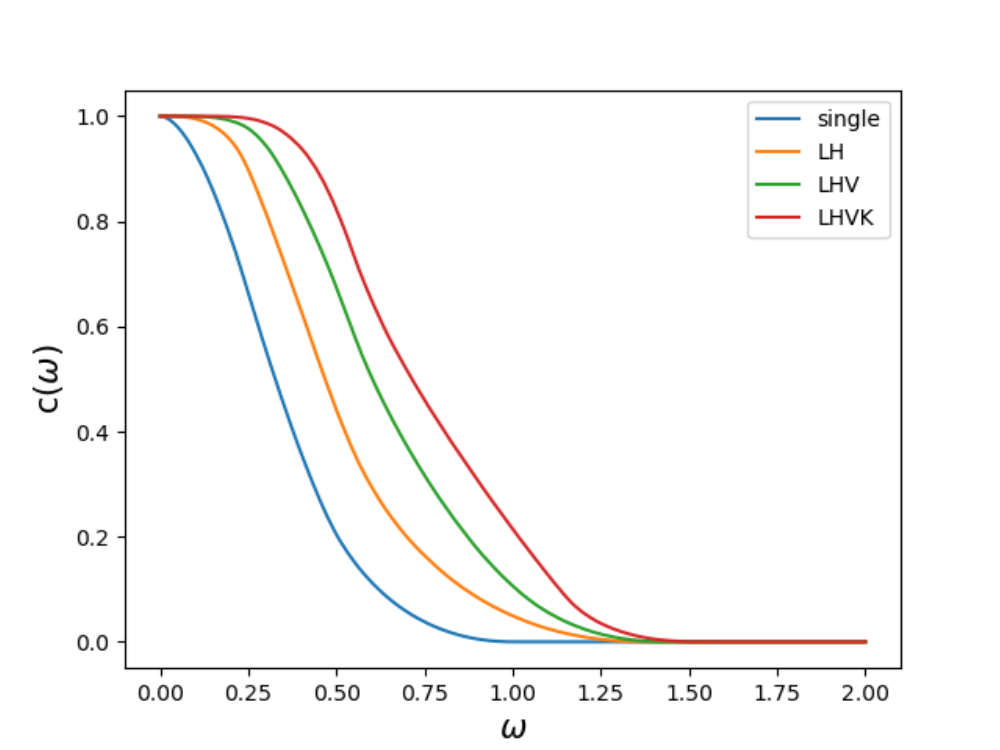}
        \caption{\label{fig:comg}Cumulative distributions of $\omega$ for a single detector, LH, LHV, and LHVK networks. }
    \end{figure}

    We calculated the cumulative distributions of $\omega$ for a single detector, LH, LHV, and LHVK via Monte Carlo simulations by randomly locating the BNS sources in the sky $(\theta, \phi)$ and their inclination $\iota$ and polarization $\psi$, as shown in Fig.~\ref{fig:comg}. 

    We compared our results with Dominik's in Appendix A of \cite{dominik2015double}, which gives numerical fits for a single and LHV detectors. They are basically consistent with ours.

\bibliography{prdtemp}

\end{document}